\documentclass[twocolumn,aps,prb]{revtex4}
\usepackage[dvipdfmx]{graphicx}
\usepackage{amssymb}
\usepackage{amsmath}
\usepackage{bm}
\usepackage{color}
\usepackage{here}
\usepackage{braket}
\def\Journal #1,#2,#3,#4#5#6#7{#1 {\bf #2}, #3 (#4#5#6#7)}

\def\Vec#1{{\bf #1}}

\begin{document}

\title{Interaction-induced insulating states in multilayer graphenes}
\author{Mikito Koshino$^{1}$, Kyoka Sugisawa$^{2}$ and Edward McCann$^{3}$}
\affiliation{$^{1}$Department of Physics, Osaka University, Toyonaka 560-0043, Japan\\
$^{2}$Department of Physics, Tohoku University, Sendai 980-8578, Japan\\
$^{3}$Department of Physics, Lancaster University, Lancaster, LA1 4YB, UK}
\date{\today}

\begin{abstract}
We explore the electronic ground states of Bernal-stacked multilayer graphenes
using the Hartree-Fock mean-field approximation and the full-parameter band model.
We find that the electron-electron interaction tends to open a band gap in multilayer graphenes from bilayer to 8-layer,
while the nature of the insulating ground state sensitively depends on the band parameter $\gamma_2$,
which is responsible for the semimetallic nature of graphite.
In 4-layer graphene,  particularly, the ground state assumes an odd-spatial-parity staggered phase
at $\gamma_2 = 0$, while an increasing, finite value of $\gamma_2$ stabilizes a different state with even parity,
where the electrons are attracted to the top layer and the bottom layer.
The two phases are topologically distinct insulating states with different Chern numbers,
and they can be distinguished by spin or valley Hall conductivity measurements.
Multilayers with more than five layers also exhibit similar ground states
with potential minima at the outermost layers,
although the opening of a gap in the spectrum as a whole is generally more difficult than in 4-layer
because of a larger number of energy bands overlapping at the Fermi energy.
\end{abstract}

\maketitle

\begin{figure}
\begin{center}
\leavevmode\includegraphics[width=1.\hsize]{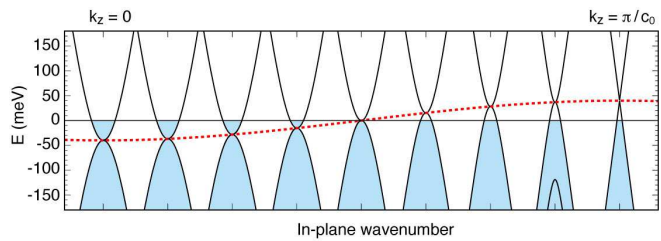}
\end{center}
\caption{Schematic band structure of graphite,
where in-plane band dispersions at various values of out-of-plane momenta $k_z$
are displayed with horizontal offsets.
The dashed curve represents the dispersion of the band
touching point, $2\gamma_2 \cos k_z c_0$, where $c_0$ is the
lattice constant in the out-of-plane direction (it is twice the interlayer spacing)
and $\gamma_2$ is a next-nearest-layer band parameter.}
\label{fig_graphite_band}
\end{figure}

\section{Introduction}

Graphite is a good conductor of electricity because of its semimetallic electronic structure,
but semimetallic behavior is not always observed as a graphitic system is thinned down to the atomic scale.
For example, in bilayer graphene, an atomically-thin film composed of only two graphene layers, \cite{mccann2013electronic,mccann2006landau,koshino2006transport,guinea2006electronic}
the electron band and the hole band with quadratic dispersions stick together at the Fermi energy
within the non-interacting model,
where the band touching point is protected by spatial inversion symmetry and time-reversal symmetry.
In experiments at low temperatures, however,
the electron-electron interaction spontaneously opens an energy gap
and the system turns into an insulator. \cite{freitag2012spontaneously,velasco2012transport,bao2012evidence,veligura2012transport,mccann2013electronic}
Electrons with different spins
tend to accumulate in different layers under the attractive exchange force,
breaking the inversion symmetry to open a gap. \cite{nilsson2006electron,min2008pseudospin,vafek2010interacting,zhang2010spontaneous,nandkishore2010quantum,zhang2011spontaneous,jung2011lattice,kharitonov2012antiferromagnetic}

The question naturally arises as to whether there is an interaction-induced energy gap
in thicker multilayer graphenes.
A recent experiment showed that an insulating state occurs at the charge neutral point 
of Bernal-stacked 4-layer graphene\cite{grushina2015insulating},
and it was followed by the observation of similar insulating behaviors
in 6-layer and 8-layer graphenes \cite{nam2016interaction}.
However, an insulating gap has not been found in Bernal-stacked
odd-layers (3-layer, 5-layer etc.) so far.
The low-energy band structure of a Bernal-stacked multilayer graphene
consists of a number of electron and hole subbands sticking near the charge neutral point:
$2M$ layer graphene is composed of  $M$ sets of bilayer-like electron and hole subbands with quadratic dispersion,
and $2M+1$ layer graphene has an extra monolayer-like linear band,
as well as $M$ sets of bilayer-like quadratic subbands.\cite{guinea2006electronic, partoens2006graphene,koshino2007orbital,koshino2008magneto,koshino2009electronic}.
Theoretically, it can be shown that all the bilayer-like subbands
can be gapped out by a staggered layer potential
$V = (-1)^{i+1}\Delta$ (where $i = 1, 2, 3, \cdots$ is a layer index
and $\Delta$ is the order parameter),
while only the monolayer-like band remains ungapped in odd-layer graphene.\cite{grushina2015insulating}
Based on this fact, it was conjectured that the insulating ground state of even-layer graphenes
would be a staggered phase in which the spin-up electrons and the spin-down
electrons alternately dominate layer by layer. \cite{grushina2015insulating}
Subsequently, the staggered phase was found to be the actual ground state in mean-field calculations
within the minimal band model. \cite{yoon2016broken}

In three-dimensional graphite, however, it is known that the band touching points of the different subbands
are not perfectly aligned but disperse along the energy axis as illustrated in Fig.\ \ref{fig_graphite_band}.\cite{koshino2009electronic}
This is actually the origin of the semimetallic property of graphite,
where an electron pocket and a hole pocket coexist at the Fermi energy.  \cite{dresselhaus2002intercalation}
The relative energy shift of the band touching point is characterized by the $\gamma_2$-parameter
which is not captured in the minimal band model,
and has a magnitude estimated to be about $20$ meV for graphite. \cite{dresselhaus2002intercalation}
If $\gamma_2$ in few-layer graphenes is of a similar magnitude as $\gamma_2$ in graphite,
we naively expect that the gap formation would be simply obstructed,
since the energy gaps in the individual subbands are masked by the relative shift of the band center.
Otherwise, there would have be some special mechanism that enables the spectral gap to overcome the semimetallic
band overlapping.

In this paper,  we explore the electronic ground states of Bernal-stacked multilayer graphenes
from bilayer to 8-layer,
using the Hartree-Fock mean-field approximation and the full-parameter band model.
We find that a small $\gamma_2$ of about 10 meV completely changes the nature of the insulating ground state,
where the system takes a different strategy to have a spectral gap under the band overlapping.
In 4-layer graphene, specifically, the ground state takes a staggered phase (odd parity)
at $\gamma_2 = 0$,  while introducing a finite $\gamma_2$
stabilizes a different state with even parity,
where the electrons are attracted to the top layer and the bottom layer,
and the energy gap opens at the crossing point of overlapping electron and hole subbands.
The even phase and odd staggered phase are topologically distinct insulating states with different Chern numbers.
Multilayers with more than 5 layers in the presence of $\gamma_2$
also exhibit similar ground states with the potential minima at the outermost layers,
although the gap opening for the whole spectrum is generally harder than in 4-layer
because of a larger number of energy bands overlapping at the Fermi energy.
In odd-layer graphenes, the presence of minor band parameters (such as $\gamma_2$)
assists in band-gap opening for the monolayer-like linear band,
which tends to blur the even-odd effect predicted in the minimal model.

The paper is organized as follows.
We introduce the effective mass band model
and the mean field treatment in Sec.\ II and III, respectively.
In Sec.\ IV, we show the calculated results for the minimal band model
neglecting all the extra band parameters including $\gamma_2$,
where the staggered phase is the ground state in any layers.
In Sec.\ V, we present the calculated results of the full parameter model,
and discuss about the $\gamma_2$-stabilized even parity phase.

\section{Effective mass model}

We describe the electronic properties of Bernal-stacked multilayer
graphene using a Slonczewski-Weiss-McClure model of graphite \cite{dresselhaus2002intercalation}.
The low energy spectrum is given by states
in the vicinity of the $K_\xi$ point in the Brillouin zone,
where $\xi = \pm 1$ is the valley index.
We define $|A_j\rangle$ and $|B_j\rangle$ as Bloch functions at the $K_{\xi}$
point, corresponding to the $A$ and $B$ sublattices of layer $j$,
respectively, where $j = 1 , \cdots , N$. In the basis of $|A_1\rangle,|B_1\rangle$;
$|A_2\rangle,|B_2\rangle$; $\cdots$; $|A_N\rangle,|B_N\rangle$,
the one-body Hamiltonian of multilayer graphene
\cite{guinea2007electronic, partoens2006graphene,koshino2007orbital,koshino2008magneto}
in the vicinity of the $K_{\xi}$ valley is
\begin{eqnarray}
H =
\begin{pmatrix}
 H_0 & V & W & & \\
 V^{\dagger} & H_0' & V^{\dagger}& W'& \\
 W & V & H_0 & V & W & \\
& W' & V^{\dagger} & H_0' & V^{\dagger} & W' & \\
  & &  \ddots & \ddots & \ddots & \ddots & \ddots
\end{pmatrix},
\label{eq_H}
\end{eqnarray}
with
\begin{eqnarray}
&& H_0 =
\begin{pmatrix}
 0 & v \pi^\dagger \\ v \pi & \Delta'
\end{pmatrix},
\quad
H_0' =
\begin{pmatrix}
 \Delta' & v \pi^\dagger \\ v \pi & 0
\end{pmatrix},
\label{Hdef}
\\
&& V =
\begin{pmatrix}
 -v_4\pi^\dagger & v_3 \pi \\ \gamma_1 & -v_4\pi^\dagger
\end{pmatrix} , \label{Vdef}
\\
&& W =
\begin{pmatrix}
 \gamma_2/2 & 0 \\ 0 & \gamma_5/2
\end{pmatrix} , \quad
 W' =
\begin{pmatrix}
 \gamma_5/2 & 0 \\ 0 & \gamma_2/2
\end{pmatrix} .
\label{Wdef}
\end{eqnarray}
Here $\pi = \hbar(\xi k_x + i k_y)$ and $\Vec{k} = -i\nabla$.
The diagonal blocks $H_0$ and $H'_0$ are intralayer Hamiltonians
for odd layer and even layers, respectively.
Also, $v = \sqrt{3} a \gamma_0/2\hbar$ is the band velocity of monolayer graphene
where $\gamma_0$ is the nearest-neighbor intralayer hopping
and $a$ is the lattice constant.
The off-diagonal matrix $V$ describes the nearest-neighbor interlayer interaction,
where $\gamma_1$ is the vertical hopping between the dimer sites (those which lie directly
above or below a site in an adjacent layer),
and the velocity parameters $v_3$ and $v_4$ are related to the oblique hopping parameters
$\gamma_3$ and $\gamma_4$ by $v_i = \sqrt{3} a \gamma_i/2\hbar$.
Matrices $W$ and $W'$ describe coupling between next-nearest
neighboring layers, and they only exist for $N \geq 3$. Parameters
$\gamma_2$ and $\gamma_5$ couple a pair of non-dimer sites and
a pair of dimer sites, respectively.
Parameter $\Delta'$ represents the energy difference
between dimer sites and non-dimer sites. It is related to the graphite band parameters as
$\Delta' = \Delta -\gamma_2 + \gamma_5$.
In the following, we adopt parameter values\cite{dresselhaus2002intercalation} $\gamma_0 = 3$ eV,
$\gamma_2 = -0.02$ eV,  $\gamma_3 = 0.3$ eV, $\gamma_4 = 0.04$ eV,
$\gamma_5 =  0.04$eV and $\Delta' = 0.05$eV.

For comparison, we also perform the same analysis
using the minimal model, where only two parameters $\gamma_0(\propto v)$ and $\gamma_1$
are retained in the band Hamiltonian above. The Hamiltonian matrix is then given by
\begin{eqnarray}
H =
\begin{pmatrix}
 H_0 & V & & & \\
 V^{\dagger} & H_0 & V^{\dagger}&  & \\
  & V & H_0 & V &  & \\
&  & V^{\dagger} & H_0 & V^{\dagger} &  & \\
  & &   & \ddots & \ddots & \ddots &
\end{pmatrix},
\label{eq_H_mini}
\end{eqnarray}
with
\begin{eqnarray}
&& H_0 =
\begin{pmatrix}
 0 & v \pi^\dagger \\ v \pi & 0
\end{pmatrix},
\quad
 V =
\begin{pmatrix}
 0 & 0 \\ \gamma_1 & 0
\end{pmatrix}.
\end{eqnarray}

\begin{figure*}
\begin{center}
\leavevmode\includegraphics[width=0.8\hsize]{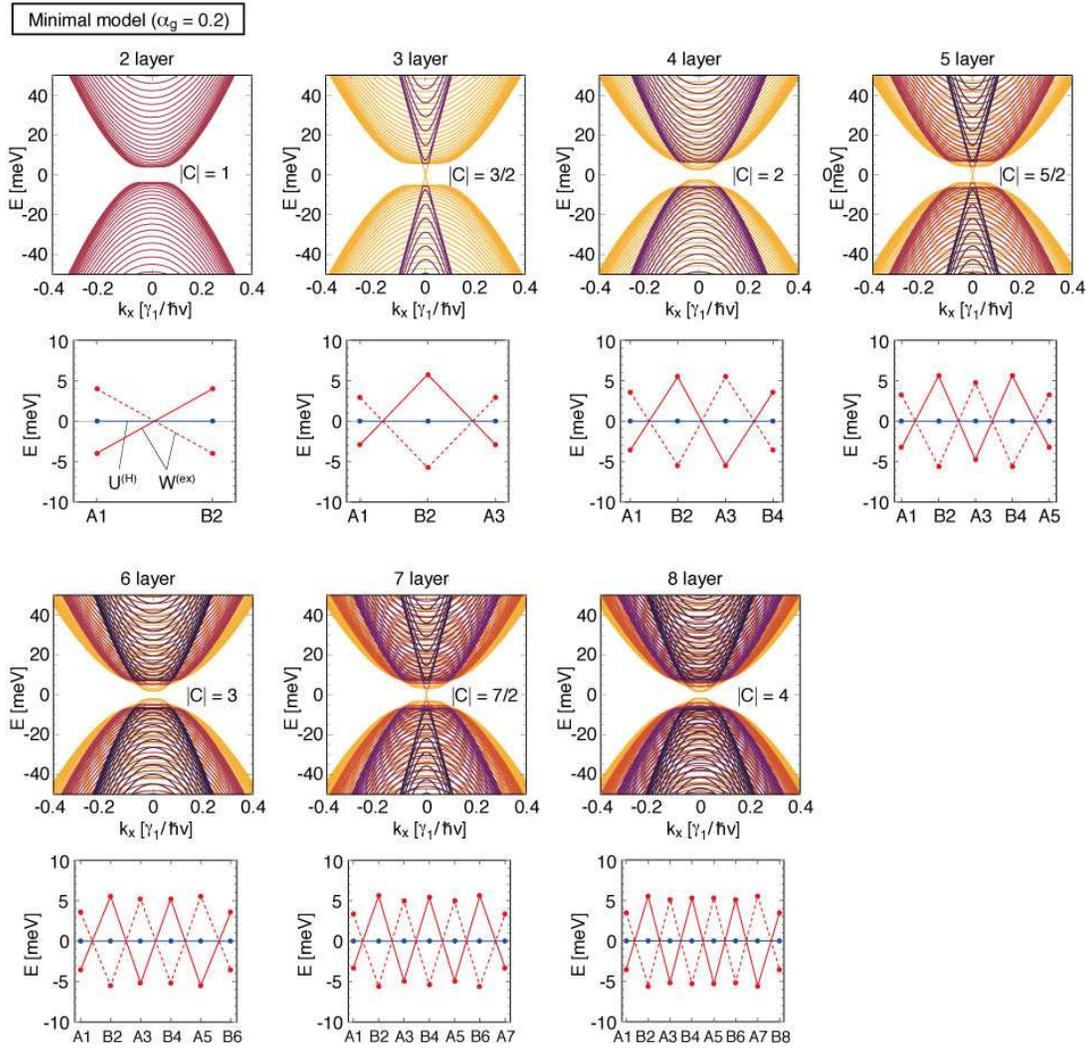}
\end{center}
\caption{Band structure of multilayer graphenes (2 layer to 8 layer)
in the minimal model with $\alpha_g = 0.2$.
The lower panels plot the Hartree potential $U^{\rm (H)}_X$ and
the exchange potential $W^{\rm (ex)}_{\Vec{k}=0\,\sigma;XX}$,
at non-dimer sites on successive layers.
}
\label{fig_minimal_model}
\end{figure*}

\begin{figure*}
\begin{center}
\leavevmode\includegraphics[width=0.95\hsize]{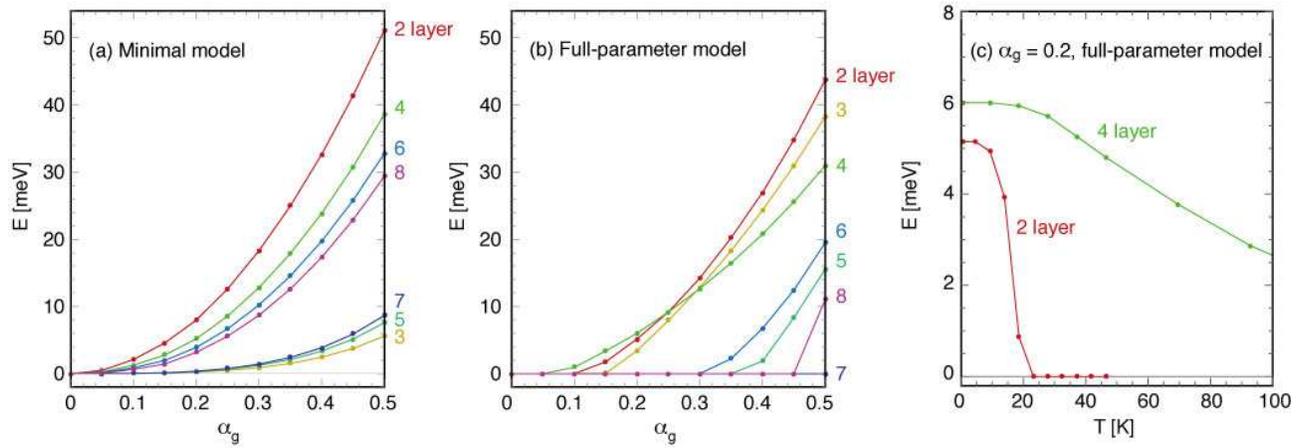}
\end{center}
\caption{Band gap against $\alpha_g$
in multilayer graphenes (2-layer to 8-layer) in (a) the minimal model
and (b) the full-parameter model.
(c) Band gap plotted against temperature
for 2-layer to 4-layer graphenes at $\alpha_g = 0.2$ in the full-parameter model.
}
\label{fig_gap_width}
\end{figure*}

\section{Mean-field theory}

We include the electron-electron interaction in the mean-field approximation.
The total Hamiltonian is written as $\hat{H}_{\rm tot} = \hat{H}+\hat{V}_{\rm MF}$ with
\begin{align}
&\hat{H}=
 \sum_{\Vec{k}\sigma XX'}
 H_{\Vec{k};XX'} c^{\dagger}_{\Vec{k}\sigma X'}c^{ }_{\Vec{k}\sigma X},
\\
&\hat{V}_{\rm MF}=
 \sum_{\Vec{k}\sigma XX'}
\left[ U^{\rm (H)}_{X}\delta_{XX'}-W^{\rm (ex)}_{\Vec{k}\sigma;XX'} \right]
 c^{\dagger}_{\Vec{k}\sigma X'}c^{ }_{\Vec{k}\sigma X},
 \label{eq_MF}
\end{align}
where $\Vec{k}=(k_x,k_y)$ is the Bloch wavenumber measured from $K_\pm$ points,
$X=A_1, B_1, A_2, B_2,\cdots$ represents the sublattice degree of freedom,
$\sigma$ is the combined index for spin ($\uparrow,\downarrow$) and valley ($K_+,K_-$),
and  $c^{\dagger}_{\Vec{k}\sigma X'}$ and $c^{ }_{\Vec{k}\sigma X}$
are the electron creation and annihilation operators, respectively.
$\hat{H}$ is the non-interacting band Hamiltonian where $H_{\Vec{k};XX'}$ is the matrix element of Eq.\ (\ref{eq_H}).
$\hat{V}_{\rm MF}$ is the interaction part, where
the first term $U^{\rm (H)}$ and second term $W^{\rm (ex)}$ are the Hartree and exchange potentials, respectively.
They are defined as
\begin{align}
&U^{\rm (H)}_{X}= \lim_{q\to 0}\sum_{X'}{v}_{XX'}\left(\Vec{q}\right)n_{X'},
\label{eq_U}\\
& n_{X}= \frac{1}{L^2}\sum_{\Vec{k}\sigma}\braket{c^\dagger_{\Vec{k}\sigma X}c^{}_{\Vec{k}\sigma X}}
-n_0,
\label{eq_n}\\
&W^{\rm (ex)}_{\Vec{k}\sigma;XX'} =
\frac{1}{L^2}\sum_{\Vec{k}\sigma}
{v}_{XX'}\left(\Vec{k}-\Vec{k}'\right)
\braket{c^\dagger_{\Vec{k}'\sigma X}c^{}_{\Vec{k}'\sigma X}}.
\label{eq_W}
\end{align}
Here $L^2$ is the system area, and
\begin{align}
{v}_{XX'}(\Vec{q}) = \frac{2\pi e^2}{\varepsilon_r q} e^{-q |z_X-z_X'|},
\end{align}
is the two-dimensional Fourier transform of the Coulomb potential,
and $z_X$ is the out-of-plane position of sublattice $X$.
Here $\varepsilon_r$ is
the dielectric constant in the interlayer spaces without the screening effect of $\pi$-band electrons,
which depends on the environment (e.g., substrate).
The summation in $\Vec{k}$ is taken within the cutoff circle $k < k_c$.
Here we take $\hbar v k_c=1$eV, which is large enough to achieve convergence
for moderate electron-electron interaction as considered below.
In Eq.\ (\ref{eq_n}), $n_0$ represents the density of the positive background charge in the system.
In the following, we consider the charge neutral case so that $n_0$ is determined by
$\sum_X n_X =0$.
Then Eq.\ (\ref{eq_U}) is reduced to
\begin{align}
&U^{\rm (H)}_{X}= -\frac{2\pi e^2}{\varepsilon_r}\sum_{X'}n_{X'}|z_X-z_X'|.
\end{align}
The label $\sigma$ takes four different configurations
$(K_+,\uparrow),(K_+,\downarrow),(K_-,\uparrow),(K_-,\downarrow)$.
Here we neglect the exchange interaction between $K_\pm$,
which corresponds to ${v}_{XX'}(\Vec{q})$ with large momentum $\Vec{q} = \Vec{K}_+-\Vec{K}_-$,
and then $\sigma$ can be treated just as a pseudo-spin label $\sigma=1,2,3,4$.

The strength of the electron-electron interaction is characterized by the effective fine structure constant
for graphene,
\begin{align}
\alpha_g = \frac{e^2}{\varepsilon_r \hbar v}.
\end{align}
As a typical value, for example, $\varepsilon_r=2$ gives $\alpha_g \sim 1$.
In the simulation, however, $\alpha_g$ should be effectively smaller than the bare value,
considering that the Hartree-Fock approximation generally overestimates the exchange interaction.
For bilayer graphene, $\alpha_g$ is supposed to be of the order of 0.1
for a quantitative agreement with experiment. \cite{jung2011lattice}
In the present study, we treat $\alpha_g$ as a parameter in the range  of $0< \alpha_g < 0.5$.

We obtain the ground state by numerically solving the above equations.
We start with an initial state with some particular configurations
for the Hartree potential $U^{\rm (H)}$ and exchange potential $W^{\rm (ex)}$,
and obtain the band structure and eigenfunctions by diagonalizing the Hamiltonian matrix.
Then we calculate the next generation of $U^{\rm (H)}$ and $W^{\rm (ex)}$ using
Eqs.\ (\ref{eq_U})-(\ref{eq_W}) with the expectation values estimated using the obtained eigenfunctions.
Using new potential terms, we again calculate new eigenfunctions,
and iterate the process until the potential terms converge.
We confirm that the final state at convergence does not depend on the choice of the initial state.
It sometimes branches into different solutions depending on the initial states,
and then we determine the real ground state by comparing the total energy of the electronic system.

\section{Minimal model}

Figure\ \ref{fig_minimal_model} shows the band structures for the minimal band model
from 2-layer to 8-layer graphenes with $\alpha_g=0.2$.
The bottom panels plot the Hartree potential $U^{\rm (H)}_X$ and
the diagonal terms of the exchange potential at the valley center, $W^{\rm (ex)}_{\Vec{k}=0\,\sigma;XX}$,
at non-dimer sites on successive layers $X=A1,B2,A3,B4,\cdots$.
For $W^{\rm (ex)}$, the constant term is subtracted so that its mean value is zero.
The solid and dotted lines in $W^{\rm (ex)}$ represent different pseudospins, say,
$\sigma=1,2$ and $\sigma=3,4$, respectively.
In all the multilayers studied here, we observe that the ground state takes a staggered arrangement
as previously predicted \cite{grushina2015insulating,yoon2016broken},
where $W^{\rm (ex)}$ alternates its sign layer by layer.
$W^{\rm (ex)}$ in $\sigma=1,2$ and $W^{\rm (ex)}$ in $\sigma=3,4$
are the positive and negative reversals of each other,
and therefore the charge modulation exactly cancels in total, resulting in
no cost for the Hartree potential.
$W^{\rm (ex)}$ of a single species has an odd parity in space inversion,
and it is a natural extension of the bilayer's ground state breaking
inversion symmetry.

The low-energy spectrum of multilayer graphene is composed of pairs of electron and hole bands
of the various band masses, and the monolayer-like linear (i.e. zero mass) band appears only in odd-layer graphenes \cite{guinea2007electronic, partoens2006graphene,koshino2007orbital,koshino2008magneto}.
In the minimal model, all the subbands are aligned on the energy axis, {\em i.e.} the conduction and valence bands exactly touch at zero energy in the absence of the electron-electron interaction.
For $\alpha_g=0.2$, we observe that a noticeable energy gap opens in even-layer graphene,
while it is nearly vanishing in odd-layer graphene due to only a tiny gap in the linear band.
We see a general tendency that the gap width becomes greater in a subband with a heavier mass
and it is the smallest in the linear (zero mass) band.
Figure\ \ref{fig_gap_width}(a) plots the band gap in the total spectrum
as a function of $\alpha_g$.
In the even-layer graphenes, the gap increases approximately $\propto \alpha_g^2$,
and the gap is smaller for thicker multilayers.
The energy gaps in odd-layer graphenes increase much more slowly than in even-layer graphenes,
and they become significant only for $\alpha_g \geq 0.3$.
We can show that, in all the multilayers, the profiles of the Hartree and exchange potential
do not depend on $\alpha_g$, except for the energy scale.

The staggered potential gives rise to a non-trivial topological property of the band structure.
In Fig.\ \ref{fig_minimal_model}, an integer appended near the energy gap in each panel
indicates the absolute value of the total Chern number $C$
summed over all the valence bands in a single pseudo-spin branch.
The Chern number corresponds to the quantized Hall conductivity in units of $e^2/h$
contributed from the pseudo-spin.
We see that $N$-layer graphene has $|C| = N/2$ in all cases,
so that the state in each pseudo-spin sector is a Chern insulator.
The sign of the Chern number depends on the sign of the staggered potential
$W^{\rm (ex)}$ (i.e., the solid line or dashed line in Fig.\ \ref{fig_minimal_model})
and also on the chirality ($K_+$ or $K_-$).
Specifically, if the signs of $W^{\rm (ex)}$ at $A1,B2,A3,B4,\cdots$
take the values $s,-s,s,-s,\cdots$ (where $s=\pm 1$)
for $K_\xi$ valley, then $C = s\xi |C|$.
The total Hall conductivity is given by the summation of $C$ over all the spin and valley sectors.
We can have various states
such as  the quantum Hall state and the quantum spin (or valley) Hall state
depending on how to assign $s=\pm 1$ to $\xi =\pm 1$ valleys,  \cite{mccann2013electronic}
while they are not energetically distinguishable in the present calculation
which treats the pseudo-spins $\sigma=1,2,3,4$ as independent sectors.

\section{Full-parameter model}

\begin{figure*}
\begin{center}
\leavevmode\includegraphics[width=0.95\hsize]{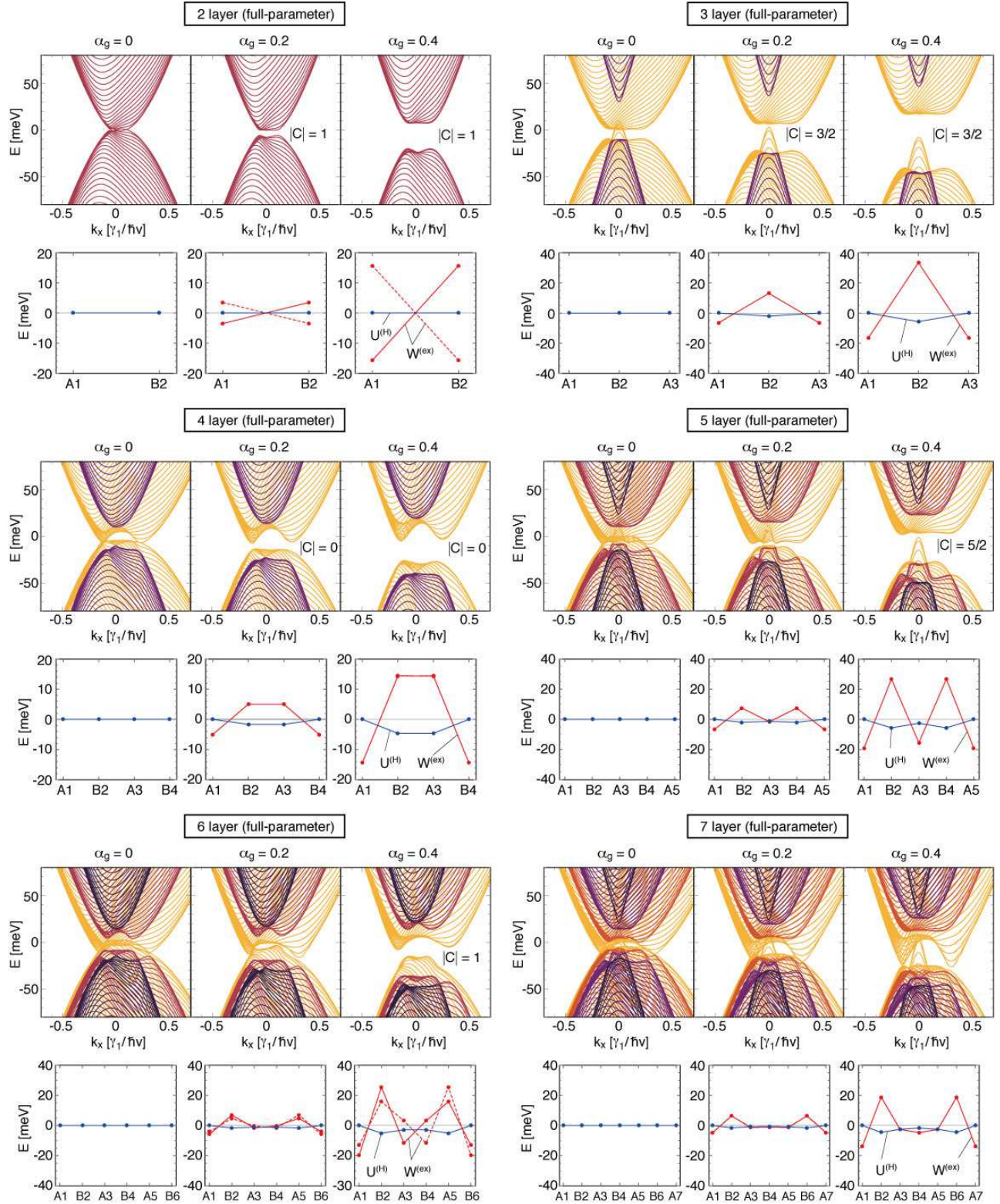}
\end{center}
\caption{Band structure of multilayer graphenes (2-layer to 7-layer) in the full parameter model,
with $\alpha_g = 0, 0.2$ and 0.4.
The lower panels plot the Hartree potential $U^{\rm (H)}_X$ and
the exchange potential $W^{\rm (ex)}_{\Vec{k}=0\,\sigma;XX}$ at non-dimer sites on successive layers.
}
\label{fig_2-7lyr}
\end{figure*}

\begin{figure}
\begin{center}
\leavevmode\includegraphics[width=1.\hsize]{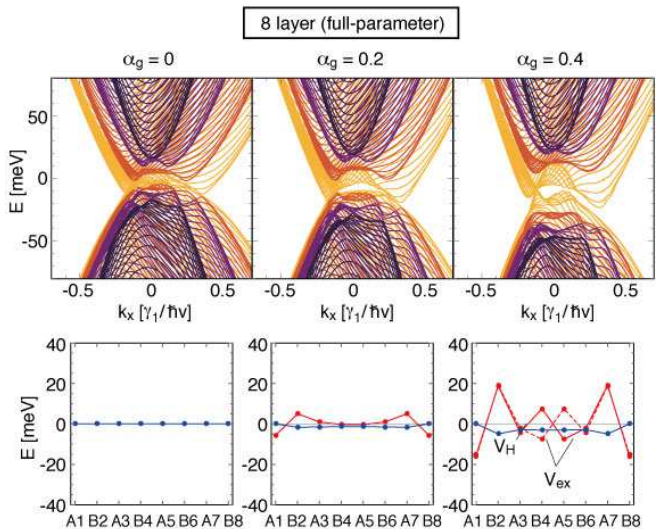}
\end{center}
\caption{Plot similar to Fig.\ \ref{fig_2-7lyr} for 8-layer graphene.
}
\label{fig_8lyr}
\end{figure}

\begin{figure*}
\begin{center}
\leavevmode\includegraphics[width=0.95\hsize]{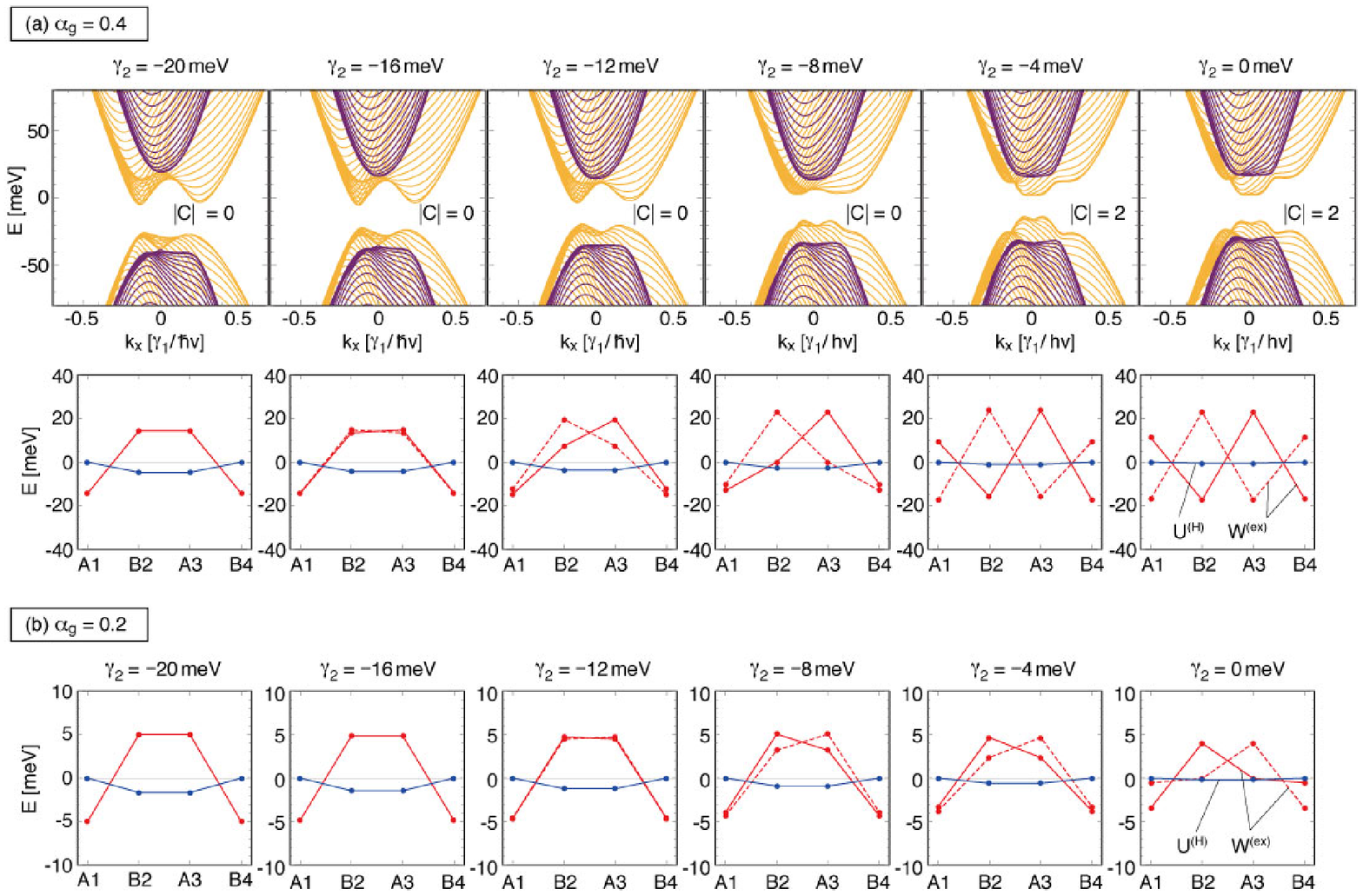}
\end{center}
\caption{(a) Band structure and the mean-field potentials (see, Fig.\ \ref{fig_2-7lyr})
of full-parameter 4-layer graphenes at $\alpha_g = 0.4$,
with different values of $\gamma_2$.
(b) Similar plots for the mean-field potentials at $\alpha_g = 0.2$.}
\label{fig_gm2}
\end{figure*}

The nature of the ground state changes drastically when additional band parameters are included.
Figure \ref{fig_2-7lyr} presents the band structures for the full parameter models
of 2 layer to 7 layer with $\alpha_g=0, 0.2$ and 0.4.
The panel of 8-layer is shown in Fig.\ \ref{fig_8lyr} .
In bilayer, the tendency is almost unchanged from the minimal model.
In 4-layer, however, the ground state takes an inversion-symmetric configuration with even parity,
in a striking contrast to the staggered phase observed in the minimal model.
Also it takes the identical configuration regardless of the pseudo-spins.
We observe similar ground states in other multilayers for $N\geq 3$,
where $W^{\rm (ex)}$ takes a low value at the outermost layers,
and its highest value at the second outer-most layers, independently of the pseudo-spin.

The appearance of a different type of ground state
originates from the presence of the next-nearest layer hopping parameter $\gamma_2$ which only
exists for $N\geq 3$.
In the full-parameter model,
the band edge of electron and hole bands are not aligned in energy in the absence of the electron-electron interaction ({\em i.e.} the conduction and valence bands do not touch exactly at zero energy)
unlike in the minimal model.
This is due to the $\gamma_2$ parameter,
which is responsible for the semi-metallic nature of graphite
where the electron Fermi surface and the hole Fermi surface coexist. \cite{dresselhaus2002intercalation}
Therefore, opening a small gap at the band touching point of each electron-hole pair
does not result in a gap opening of the total spectrum,
so that the system needs to adopt a different strategy in order to open a gap.
Actually, we can show that the outermost attractive potential observed here
partially cancels the effect of $\gamma_2$ and aligns the band centers (See, appendix).
This assists the opening of an energy gap by reducing the band overlapping at the Fermi energy.

In Fig.\ \ref{fig_gm2}(a), we present the band structure and the mean-field potential
in the full-parameter model for 4-layer graphene with $\alpha_g=0.4$,
in which the $\gamma_2$ parameter is reduced from its value in graphite ($-20$ meV) down to zero
with all other parameters unchanged.
We see that the ground state gradually changes from the even-parity phase to the odd-parity, staggered phase
as $\gamma_2$ decreases.
An interesting observation is that the total Chern number below the gap, $C$, is zero in the even phase
while it is 2 in the staggered phase, so there is a topological phase transition at some stage as $\gamma_2$ decreases.
$C$ vanishes in the even phase, because the surviving inversion symmetry
and time-reversal symmetry
requires a vanishing Berry phase everywhere except at the band-degeneracy point. \cite{haldane2004berry,koshino2013electronic}
Experimentally, the non-zero Hall conductivity in a single pseudo-spin sector causes a valley or spin Hall effect, \cite{mccann2013electronic}
so that the even and odd phases can be distinguished by measuring it.
In Fig.\ \ref{fig_gm2}(b), we show similar plots of the mean-field potential for weaker interaction, $\alpha_g=0.2$.
We observe that the even phase survives down to smaller $\gamma_2$
compared to $\alpha_g=0.4$,
suggesting that the $\gamma_2$-stabilized even phase is more stable for weaker interactions.

Generally, the band touching points of the intrinsic even-layer graphene are protected by spatial
inversion symmetry and time-reversal symmetry. \cite{manes2007existence,koshino2013electronic}
More specifically, the coexistence of the two symmetries
requires the vanishing of the Berry curvature at any non-degenerate band states,
and this immediately concludes that a band touching point with non-trivial Berry phase cannot be split,
because otherwise non-zero Berry curvature arises where the degeneracy is lifted.
In 4-layer graphene, the band gap opening in the even-parity state may appear to contradict with this fact,
but, there, band touching actually remains between the first and the second bands
and also between the third and the fourth bands (out of the four low-energy bands)
so a band gap can open at the center without breaking these degenerate points.
In bilayer graphene, on the other hand, there are only two bands at the charge neutral point,
so that inversion-symmetry breaking is the only way to open a gap,
as actually observed in Fig.\ \ref{fig_2-7lyr}.
The 6-layer graphene has a similar situation to bilayer where the six low-lying bands touch in pairs of (1,2), (3,4) and (5,6)
under the inversion symmetry, and the middle pair (3,4) prevents a gap opening at the charge neutral point.
In the numerical calculation shown in Fig.\ \ref{fig_2-7lyr},
we actually see that the gapped state at $\alpha_g = 0.4$
slightly breaks the inversion symmetry (of $W^{\rm (ex)}$ in each single species)
on top of the even-parity feature.
The above argument for band touching
does not apply to odd-layer graphenes where the inversion symmetry is intrinsically absent.

The energy gap width as a function of $\alpha_g$ is shown in Fig.\ \ref{fig_gap_width}(b).
For $\alpha_g \leq 0.2$, which is supposed to match the real experimental situation,
the energy gap is the largest in 4-layer, in which it is even bigger than in bilayer.
Thicker graphenes ($N \geq 5$) require a large interaction $\alpha_g > 0.3$ to have a gap
because of the severely overlapping energy bands (in the absence of interactions).
However, 3-layer graphene is atypical because its gap is much more significant for $\alpha_g > 0.15$ than in the minimal model.
In the full parameter model at $\alpha_g=0$,
the additional band parameters $\gamma_2$, $\gamma_5$ and $\Delta$ cause the monolayer and bilayer-like bands in 3-layer to overlap each other (hence there is no overall band gap), but, when viewed separately, the monolayer bands are gapped and the bilayer-like bands are gapped, too. This then assists the gap opening of the whole spectrum for $\alpha_g \neq 0$.

Figure \ref{fig_gap_width}(c) shows the temperature dependence of the energy gap
of 2- and 4-layer graphenes at $\alpha_g=0.2$.
It shows that the gap in 4-layer remains up to much higher temperature than in bilayer,
even though at zero temperature they are of a similar order.
The critical temperature for the gap closing is determined by
the characteristic energy scale of the band structure contributing to the total energy minimization.
In the bilayer, the change of the energy band due to the electron-electron interaction
only takes place near the gap within the energy scale about a few meV at $\alpha_g=0.2$
as seen in Fig.\ \ref{fig_2-7lyr}.
In 4-layer, on the other hand, the band structure near $k=0$ already has some complex structure
in the energy range about 20~meV, and they all contribute to the charge transfer and the total energy reduction.
To smear out the gap, we need a temperature to match this energy scale, and it becomes much greater than
in bilayer.

\section{Conclusion}

We studied the electronic band structure in Bernal-stacked multilayer graphenes under
the electron-electron interaction, using
the mean-field theory and the band model fully including band parameters.
We demonstrated that the ground state is governed by the semimetallic parameter $\gamma_2$,
where the previously-conjectured staggered state yields to an even-parity state
in the presence of $\gamma_2$, where the outermost layers have the lowest potential energy.
We also found that the staggered phase and the even-parity phase have different Chern numbers
and they can be distinguished by valley / spin Hall conductivity measurements.

In the ground state calculation,
we treat $\gamma_2$ and the interaction strength $\alpha_g$ as variable parameters,
and quantitatively estimate the energy gap width in changing parameters.
Experimentally, the insulating gap was observed not only in Bernal-stacked bilayer \cite{freitag2012spontaneously,velasco2012transport,bao2012evidence,veligura2012transport,mccann2013electronic},
but also in 4-layer\cite{grushina2015insulating}, 6-layer and 8-layer graphenes\cite{nam2016interaction}.
According to Fig.\ \ref{fig_gap_width}, the opening of a gap for 6 and 8-layers within the full-parameter model
requires a fairly high interaction $\alpha_g \agt 0.4$,
but such a high interaction would result in bilayer's gap being more than 30 meV,
which is inconsistent with experiment.
This suggests that the value of $\gamma_2$ in few-layer graphenes may be actually lower than in graphite,
and the reality could be somewhere between  Figs.\ \ref{fig_gap_width}(a) and (b).
While the reason for suppression of $\gamma_2$ is unclear,
it is possible, for example, that the optimum lattice structure of a few-layer device
takes a different interlayer spacing from graphite's, and that modifies $\gamma_2$.\cite{nam2016interaction}
The $\gamma_2$ is a small band parameter corresponding to the next-nearest neighbor interlayer hopping,
and it should be sensitive to the interlayer spacing.
Unknown values of band parameters including $\gamma_2$
should be clarified by detailed spectroscopic measurements for few-layer graphenes.
It is also conceivable that other factors,
such as the effect of disorder or of interactions beyond mean-field theory, would be significant
for further quantitative arguments. The study of these issues is left as an open question.

\begin{acknowledgments}
The authors thank Alberto F. Morpurgo and Kazuhiko Kuroki for helpful discussions.
This work was supported by JSPS KAKENHI Grant Numbers  JP15K21722, JP25107001 and JP25107005.
\end{acknowledgments}

\bibliography{mlg_hf}


\end{document}